# Evidence for a New Intermediate Phase in a Strongly Correlated 2D System near Wigner Crystallization

R. L.J. Qiu[1], N. J. Goble[1], A. Serafin[2], L. Yin[2], J. S. Xia[2], N. S. Sullivan[2], L. N. Pfeiffer[3], K. W. West[3], and X. P.A. Gao[1], *

[1]Department of Physics, Case Western Reserve University, Cleveland, Ohio 44106, USA.
[2] National High Magnetic Field Laboratory and Department of Physics, University of Florida, Gainesville, Florida 32611, USA.
[3]Department of Electrical Engineering, Princeton University, Princeton, New Jersey 08544, USA.

*Email: xuan.gao@case.edu

**Abstract**
How the two dimensional (2D) quantum Wigner crystal (WC) transforms into the metallic liquid phase remains to be an outstanding problem in physics. In theories considering the 2D WC to liquid transition in the clean limit, it was suggested that a number of intermediate phases might exist. We have studied the transformation between the metallic fluid phase and the low magnetic field reentrant insulating phase (RIP) which was interpreted as due to WC formation [Qiu et al, PRL **108**, 106404 (2012)], in a strongly correlated 2D hole system with large interaction parameter $r_s$ (~20-30) and high mobility. Instead of a sharp transition, we found that increasing density (or lowering $r_s$) drives the RIP into a state where the incipient RIP coexists with Fermi liquid. This apparent mixture phase intermediate between Fermi liquid and WC also exhibits a non-trivial temperature dependent resistivity behavior which may be qualitatively understood by the reversed melting of WC in analogy to the Pomeranchuk effect in the solid-liquid mixture of Helium-3.

PACS numbers: 67.10.Jn , 71.27.+a, 71.30.+h, 73.20.Qt, 73.40.c,

**Main Text:**

Understanding different phases of matter and transitions between them has been a rich subject in science. Modern research has revealed many intriguing phenomena when low dimensionality, Coulomb interactions, or disorder become prominent in phase transitions. In 2D, it has been known that phases with long range order are prohibited[1]. Nevertheless, quasi-long range ordered state at low temperature ($T$) is possible, and raising the temperature above a critical value can drive the system into a disordered state according to Berezinskii-Kosterlitz-Thouless (BKT) theory[2,3]. Unlike conventional second order phase transitions given by the Landau theory, a BKT transition is due to the binding/unbinding of topological defects (vortices and anti-vortices)[2,3]. BKT successfully describes a number of important phase transitions in condensed matter such as the 2D superconductivity and superfluid transitions, the 2D XY model, and the first order crystal-liquid transition in 2D[4].

The melting of a 2D electronic crystal (or WC[5]), has been an unresolved problem for long time. Early work by Halperin, Nelson[6] and Young[7] predicted that the low $T$ 2D WC melts into an intermediate anisotropic fluid phase (hexatic 'liquid crystal') with quasi-long-range bond orientation order but without positional order, before becoming a disordered isotropic fluid at high $T$. Considering the density driven melting of a 2D WC at low $T$ and assuming a direct first order transition from crystal to fluid, variational Monte-Carlo simulations obtained a critical value for the interaction parameter $r_s$, the ratio between Coulomb and kinetic energy, to be ~37(ref.8). Specifically, in 2D, $r_s$ is given by $r_s = 1/(a^*\sqrt{\pi p})$ with $p$ and $a^* = \hbar^2\varepsilon/(m^*e^2)$ being the carrier density and effective Bohr radius where $m^*$, e, $\hbar$, $\varepsilon$ are the effective mass of carriers, electron charge, reduced Planck's constant, and dielectric constant. However, further theoretical calculations suggested that other complex quantum phases may exist between the 2D WC and Fermi liquid, at $1<<r_s<37$[9-14]. More generally, Jamei, Kivelson and Spivak presented a theorem concluding that the direct first order 2D WC-liquid transition is prohibited in the presence of Coulomb interactions, thus phase separation at mesoscopic scales and new intermediate phases (electronic 'micro-emulsions') are unavoidable as a result of Coulomb frustration[15]. A conceptually simple example of intermediate micro-emulsions is short-range ordered bubbles/stripes of WC coexisting with Fermi fluid at a mesoscopic scale which is predicted to have interesting thermodynamic and transport properties analogous to the mixture of Helium-3 solid and fluid[10, 11, 16]. Such Coulomb-frustrated electronic phase separation is believed to have broad implications in other strongly correlated materials as well (cuprate high temperature superconductors, colossal magnetoresistance manganates, etc.)[17, 18].

Despite these important theoretical progresses on the 2D melting of clean quantum WC, experimental studies of 2D WC-fluid transition are limited, and there is no direct experimental evidence of interaction-driven intermediate phases to date. The classical version of a 2D WC was realized in electrons confined on the surface of liquid Helium[19]. The realization of a 2D quantum WC has been actively sought in electron or hole carrier systems confined in semiconductor hetero-interfaces or quantum wells[20]. To reach the clean 2D WC phase in zero magnetic field ($B$=0) where $r_s$ is very large (>~37), ultra low carrier density and extremely high sample quality (purity) are needed so the interactions dominate over kinetic and disorder potential energies. Although there are sound reasons to believe that the 2D metal-insulator-transition and metallic transport in $B$=0 in a number of correlated 2D systems with high $r_s$ are related to the physics of 2D WC-liquid transition[16], 2D WC-liquid transition and intermediate phase formation are not well established in experiment and the role of disorder in this problem remains elusive. However, for 2D electrons with not so large $r_s$ (e.g. $r_s$~1-10), intense perpendicular magnetic field can help the WC formation by forcing electrons into the same Landau levels (LLs), which quenches the kinetic energy. In the extreme quantum limit where only the lowest LL is partially filled (filling factor ν<1), the ground state is in competition between the fractional quantum Hall (FQH) fluid and disorder pinned WC, which is manifested as the reentrant insulating phase (RIP) between the ν=1/3 or 1/5 FQH and the ν=1 integer QH in GaAs/AlGaAs[21,22] or the high $B$ insulating phase at the smallest ν[20, 23]. In Si metal-oxide-semiconductor-field-effect transistors, pinned 2D electron crystals were found where $r_s$ is much smaller than the critical $r_s$ for clean WC-liquid transition, presumably due to a dominant role of disorder[24, 25]. Low frequency transport[26] and microwave absorption[27] experiments on the high $B$ insulating phase at fractional filling factors revealed signatures of pinned WC melting at elevated temperatures. Moreover, the observation of anomalous low $T$ magneto-resistance oscillations in 2D holes near the vicinity of 1/3 filling factor was interpreted as mixing of fractional quantum Hall liquid and a disorder pinned solid phase by Csáthy et al[28]. However, whether the WC-liquid transition is a direct transition and if interaction driven intermediate phases exist in the clean limit still remains a mystery, due to the intricate role of disorder [29, 30] in previous experiments[20-28].

In this study, we investigate the melting of 2D WC in weak magnetic fields employing a dilute 2D hole system (2DHS) with large $r_s$ ($r_s$~20-30) residing in high mobility GaAs quantum wells (QWs) with 10nm width. The strongly interacting character of the dilute 2DHS and high sample quality has enabled us [31] to observe the RIP due to WC formation at low $B$, which is connected to the $B$=0 metal-insulator transition at $r_s$~37, approaching the clean limit 2D WC-liquid transition phase diagram as expected in theory [32]. Here we examine the transport properties of this low $B$ WC state over extended temperature and density ranges to understand the melting of 2D WC. Our data show that when disorder is weak, the 2D WC-liquid transition gives way to phase separation and forms a new intermediate phase where WC coexists with metallic fluid. This intermediate phase tends to freeze upon raising temperature, a bizarre quantum phenomenon similar to the Pomeranchuk effect [33] found in the mixture of liquid and solid Helium-3, as the micro-emulsion theory suggested[10, 11, 16].

Low frequency magneto-transport measurements using lockin technique were primarily performed on four high mobility GaAs QW samples from two wafers (wafer #5 with growth sequence number 5-24-01.1, and wafer #11 with growth sequence number 11-29-10.1) grown on (311)A GaAs similar to previous work [31, 34]. Unless specifically noted, the sample studied has rectangle Hall bar shape with length ~8-9mm and width ~2-3mm and the measurement current was applied along the high mobility direction $[\bar{2}33]$. The samples have $Al_{0.1}Ga_{0.9}As$ barriers and Si delta doping layers placed symmetrically at a distance of 195 nm away from the 10 nm thick GaAs QW. Wafer #5 and #11 have density $p$~ 1.3 ×$10^{10}$/cm$^2$ and mobility μ=5 and 2×$10^5$ cm$^2$/Vs without gating. Back gate was placed approximately 150-300 μm underneath the sample and used to tune the hole density. Similar data were obtained in independent experiments covering temperature from $T$=1K down to $T$=10 mK in three different dilution refrigerators. An additional GaAs QW sample (wafer #1 with growth sequence number 1-10.11.1) with the same growth structure but lower mobility (μ=0.8×$10^5$ cm$^2$/Vs at $p$=1.3 ×$10^{10}$/cm$^2$) than wafer #5 and #11 was studied to shed light on the effects of increased disorder on the intermediate phase formation.

Figure 1a shows a qualitative sketch of the clean 2D WC-liquid transition phase diagram in the hole density ($p$) – perpendicular magnetic field ($B$) plane obtained by quantum Monte-Carlo simulations[32]. The white dashed line marks the critical boundary assuming the direct first order transition from crystal to liquid. Taking a horizontal line cut from the phase diagram, if $p$ is high (or $r_s$ is low), one expects to observe transitions from low $B$ liquid to the RIP (or WC) at 1/3<ν<1 followed by the ν=1/3 FQH liquid and then another insulating phase induced by WC at the highest $B$. This scenario was the subject of extensive study in the 90's[20-23] and confirmed in our own magneto-resistivity measurements as shown in the top panel of Fig.1B for $p$=2.25 (unit for hole density is $10^{10}$/cm$^2$ throughout this paper) in a sample from the highest mobility wafer #5. As $p$ decreases (or $r_s$ increases), increased Coulomb interaction strength favors the WC state more. As a result, the ν=1/3 FQH is taken over by the WC at high field. At even lower $p$ and with $r_s$ approaching 37, magnetic field enhanced Coulomb correlations induce a WC emerging between the $B$=0 liquid and ν=1 QH liquid, as represented by the exemplary $p$=0.86 data in the bottom panel of Fig.1b. Note that this phase diagram gives a qualitative picture of interaction driven 2D WC-liquid melting in the clean limit (without disorder). The confirmation of this phase diagram in our earlier work[31] attests to the suitability of using clean 2DHS for the study of WC-liquid melting near zero $B$ with minimal disorder effects[8-16].

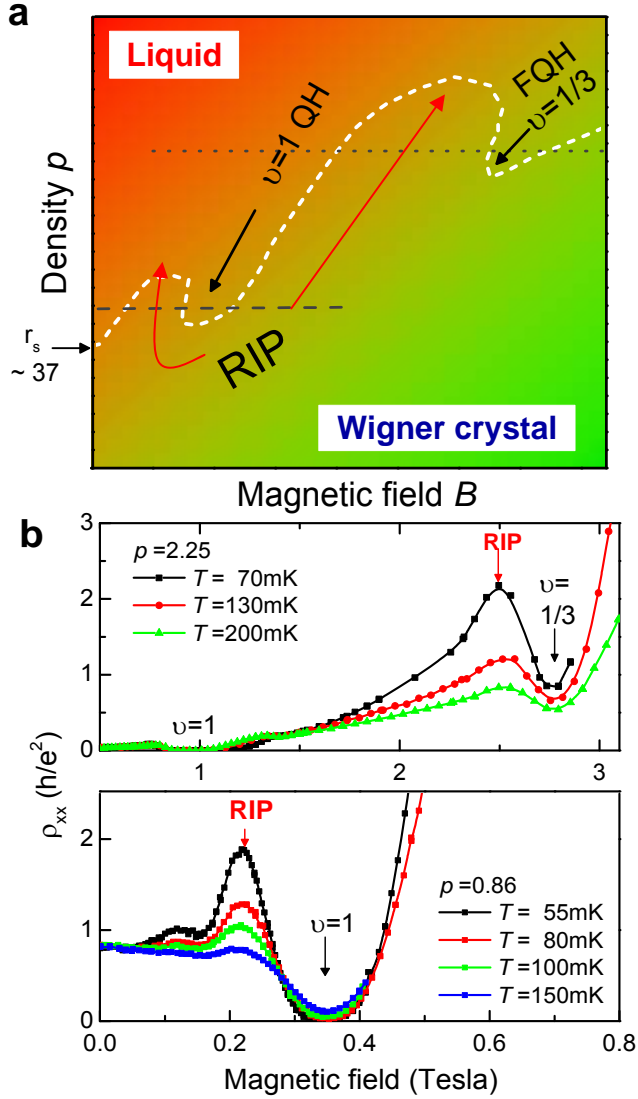

**FIG 1. a,** Schematic phase diagram of 2D Wigner crystal-liquid transition in the clean limit. Two horizontal lines (top to bottom) roughly represent the corresponding trajectories of the two field scans (top to bottom) in Fig.1b. **b,** Magneto-resistivity vs. magnetic field ($B$) of a high mobility QW (sample #5-B) for densities $p$=2.25 and 0.86 at various temperatures from 55mK to 200mK. The reentrant insulating phase (RIP) due to Wigner crystal formation is seen before the $\nu = 1/3$ FQH and $\nu = 1$ QH states as $p$ decreases.

We first investigate, at low $T$ and low $B$, how the WC merges into the fluid phase as $p$ increases. When $p$ increases, $r_s$ decreases and weakened interaction effects drive the system towards the liquid state. Instead of a sharp transition from the insulating WC into a metallic liquid at a well defined critical density, we observe the (WC induced) RIP transforms into a state characteristic of both an insulating crystal and metallic Fermi liquid. Figure 2a shows a series of

$\rho_{xx}(B)$ traces at $T$=55mK from $p$=0.86 to 1.46. At $p$=0.86, the system's $r_s$~30 is not yet large enough to fully crystallize the system at $B$=0. With the application of a small magnetic field, the 2DHS's kinetic energy is quickly quenched. The reduced wavefunction overlapping stabilizes the WC as a RIP peak near 0.23T, which has an insulating temperature dependent resistivity with a thermally activated form (Fig.3a and Ref.31). As $p$ increases, the RIP, a signature of WC, does not disappear immediately when the system develops low resistivity and clear Shubnikov de-Haas (SdH) oscillations characteristic of a Fermi fluid. Instead, the RIP peak in $\rho_{xx}(B)$ evolves into a spike inside the $v$=2 SdH dip at $B$~0.27T for $p$=1.25, 1.35 and is eventually barely detected at $p$=1.46. Figure 2b shows the $\rho_{xx}(B)$ curves from $T$=55mK to 150mK for $p$=1.25, 1.35 and 1.46 to illustrate the effect of raising temperature for the same density. We find that the incipient RIP or WC formation near 0.27T (marked by the red arrow) becomes stronger at lower temperature as indicated by the sharper spike, despite the overall metallic temperature dependence and very low resistivity value (<0.1h/e$^2$). The persistence of softened RIP on top of the regular SdH oscillations deep inside the metallic regime in Fig.2 suggests the incipient WC formation inside the metallic liquid, or *the existence of an intermediate state where WC is mixed together with a metallic Fermi fluid background*. Furthermore, this 2D WC melting induced intermediate state appears to exist over a quite broad range of resistivity value (from <0.1h/e$^2$ to ~h/e$^2$). Similar behavior was obtained in three other samples from both the highest mobility wafer #5 and second highest mobility wafer #11, and measured under both square Van der Pauw and rectangular Hall bar configurations (Fig.S1, S2). The transverse Hall resistivity shows a reduced value compared to the classical value (Fig.S3 and Ref.34), perhaps reflecting the inertial Hall response of WC components in the intermediate state.

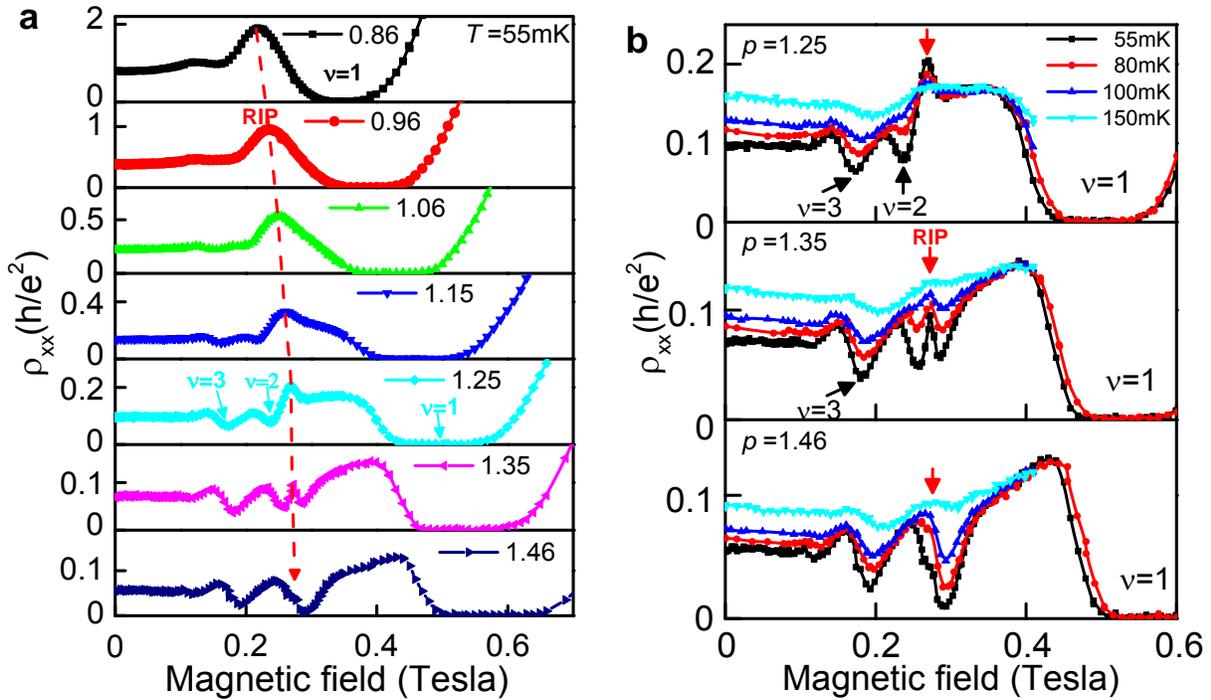

**FIG. 2. a,** Magneto-resistivity of a high mobility QW (sample #5-B) at $T$=55mK for densities $p$= 0.86-1.46 ($r_s$=30-21). The dash line illustrates the evolution of the 2D Wigner crystal or reentrant insulating phase (RIP) while increasing the density $p$. **b**, Magneto-resistivity

of sample for metallic densities $p$= 1.25, 1.35 and 1.46 at temperatures 55mK, 80mK, 100mK and 150mK. An incipient RIP is clearly seen at ~0.28T (marked by a red arrow).

Next we consider the melting of the WC due to thermal fluctuations. The maximal resistivity at the peak of RIP, $\rho_{peak}$, is plotted in a log scale against $1/T$ in Fig.3a for the densities shown in Fig.2a. The corresponding $\rho_{peak}$ vs. $T$ in linear scale is shown as Fig.3b. It is anticipated that the $\rho(T)$ of the WC follows the thermal activation law $\rho(T) \propto \exp(\Delta/2k_BT)$, with the activation energy $\Delta$ representing the energy required to create a vacancy or defect in the WC to allow charge transport. Another important temperature scale is the melting temperature $T_{melt}$, above which the WC no longer has quasi-long range crystalline order. Indeed, $\rho_{peak}(T)$ follows the thermal activation law at the lowest temperatures, indicated by the dashed lines in Fig.3a. Meanwhile, as $T$ increases, there is a fairly well defined temperature above which the thermal activation ceases. This feature was understood as the thermal melting of the 2D WC in a prior work at filling factor $\nu<1/5$[26]. Thus we determine $T_{melt}$ ~0.3K at $p$=0.86 and find it decreasing with increasing $p$, which is reasonable as the WC shall be harder to form at smaller $r_s$. At $p$=1.35, in spite of the existence of RIP spike in $\rho_{xx}(B)$ (Fig.2a), $T_{melt}$ cannot be detected down to 55mK, the base temperature of this cool down, showing that the system remains to be a liquid percolating through WC islands in the ground state, or the quasi-long range WC ordering happens at much lower temperature. What is unique in the present study where we find the intermediate state in which WC is mixed with the low resistivity metallic liquid, is the temperature dependence of $\rho_{peak}$ above $T_{melt}$. At very high temperatures (e.g. ~1K), all the densities in Fig.3a show similar temperature dependent $\rho$, due to the fact that the system has completely melted into a semi-quantum fluid at $T$ higher than the Fermi temperature, $T_F$[11,16]. Strikingly, for the highest densities, the system exhibits an intriguing reversed trend in $\rho(T)$ between the high $T$ semi-quantum fluid regime and the low $T$ WC regime, as highlighted in the shaded area in Fig.3c. Since this temperature regime is above the melting temperature of WC and below the Fermi temperature inside the intermediate state, a natural explanation for Fig.3c is the reversed melting of a WC/Fermi fluid mixture, a theoretical scenario proposed in the micro-emulsion model[10,11,16] in analogy to the Pomeranchuk effect[33] in the Helium-3 solid/liquid mixture: at intermediate temperatures ($T<T_F$), WC with unordered spins has spin entropy $S_{WC} \sim p \times \ln2$ which is larger than the entropy of a degenerate Fermi fluid ($S_{FL} \sim p \times T/T_F$), thus the WC bubbles would melt into Fermi liquid as the mixture cools, giving rise to the metallic resistivity. Only when the temperature is low enough to induce spin ordering in the WC and quench the spin entropy, the system will crystallize upon cooling like normal liquid/solid mixture and exhibit thermally activated insulating behavior. This low temperature behavior was also observed in our experiment at $T<T_{melt}$. Using $T_0$, the maxima position of $\rho_{peak}(1/T)$ to define the characteristic upper bound temperature of the intermediate crystal/fluid mixture state, we obtain Fig.3d, a density-temperature diagram for the melting of 2D WC. Because of the lack of a metallic downturn in Fig.3a for the top three traces, it is difficult to precisely define $T_0$ at the lowest densities. Even though, we see that $T_0$ tends to merge with $T_{melt}$ in the low density limit as suggested by Fig.3a and c. Figure 3d presents the following picture of the 2D WC melting: at the lowest $T$, the system is dominated by the crystal phase while the intermediate crystal/liquid mixture state dominates between $T_0$ and $T_{melt}$, with a reserved melting similar to the Pomeranchuk effect. Finally the system melts into a semi-quantum fluid at the high $T$ regime [35]. Note that data from both the highest mobility wafer (#5) and medium-high mobility wafer (#11) are included in Fig.3d. It is striking that despite the more than two times mobility difference, these two wafers show very consistent $T_0$ and $T_{melt}$, indicating that this phase diagram is intrinsically controlled by the interaction and Fermi energies, not disorder.

At the peak of RIP, one expects that the WC becomes the most robust, and tuning magnetic field towards zero or the $\nu=1$ QH state shall also cause the WC to melt. Surprisingly, we find again that the melting temperature of the WC is relatively insensitive to $B$, but the activation energy $\Delta$ varies strongly with $B$ as shown in Fig.3f, which is obtained by fitting the $\rho(1/T)$ data in Fig.3e for $p=0.96$ in various $B$. This indicates some subtle difference between the magnetic field tuned WC melting and density tuned melting, since we find that the activation energy follows the trend of $T_{melt}$ for the case of density driven melting (Fig.S4), in contrast to the $B$ tuning effect. We speculate that the melting temperature is only determined by the bare $r_s$ and interaction energy scales, which are determined by the carrier density. However, the magnetic field controls the extent of wavefunction overlapping for neighboring particles and thus is more important to the single vacancy/defect creation (and $\Delta$) in WC.

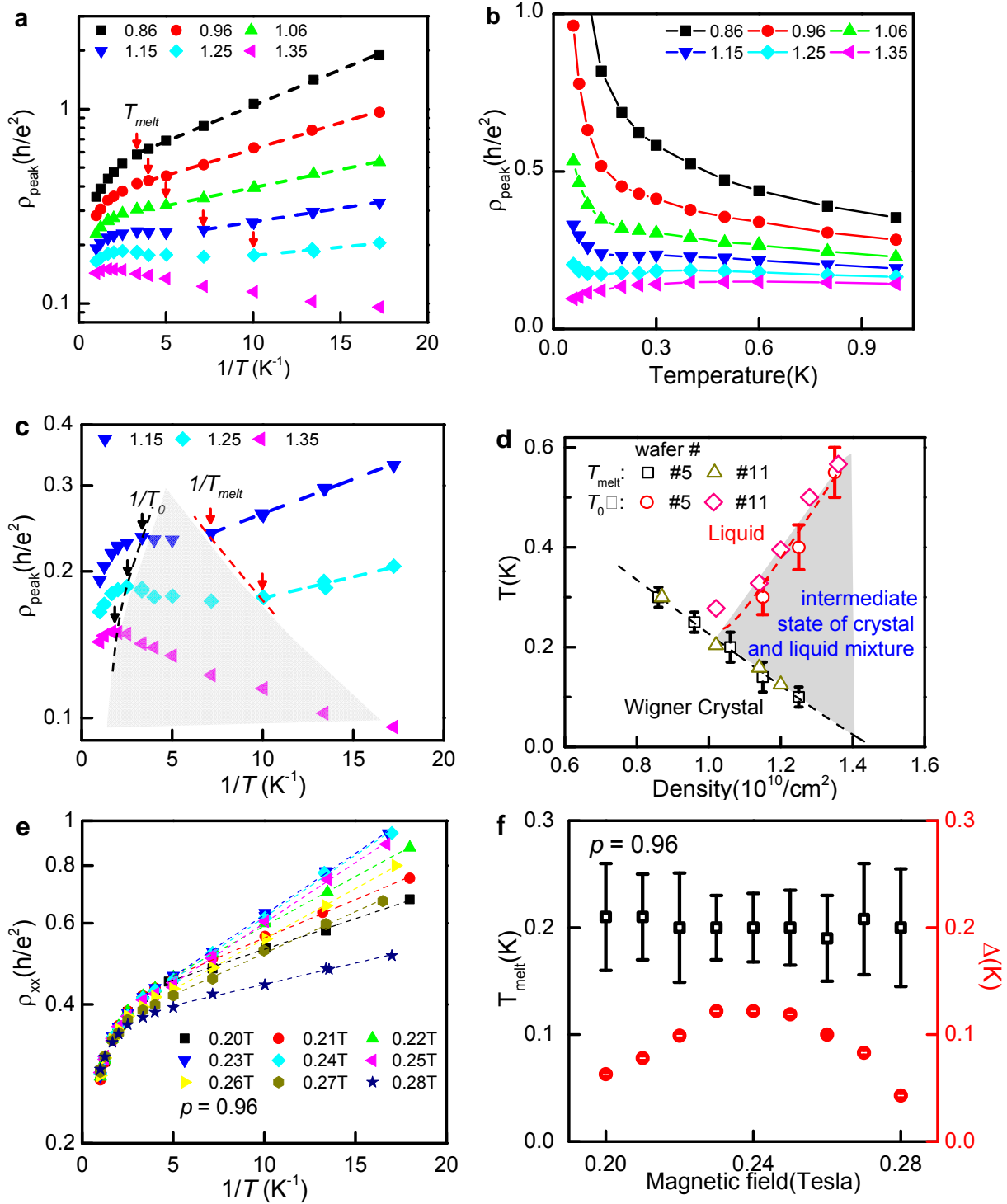

**Fig. 3.** a, Temperature dependence of resistivity at the peak of RIP from QW sample #5-B plotted vs. $1/T$ scale at densities $p$ = 0.86, 0.96, 1.06, 1.15, 1.25, and 1.35 (corresponding magnetic fields are 0.220, 0.235, 0.250, 0.260, 0.268, 0.272, and 0.274Tesla). The dash lines are the fits to activation behavior $\rho \propto \exp(\Delta/2T)$. The melting temperature $T_{\text{melt}}$ of WC is determined as the point when activated $\rho(T)$ terminates as marked by red arrows. **b**, Data in (a) plotted in linear scale. **c,** Two-step melting of 2D WC at $T_{\text{melt}}$ and $T_0$. The system shows an anomalous

metallic behavior above $T_{melt}$ but below the semi-degenerate crossover temperature $T_0$ due to the reversed melting of WC in the intermediate mixture state. **d**, Phase diagram of 2D WC-liquid transition in the density-temperature plane. **e**, Temperature dependence of resistivity for $p = 0.96$ for sample #5-B at magnetic fields $B$ = 0.2-0.28 T, from which the WC melting temperature and activation energy $\Delta$ (in units of K) are extracted and plotted as a function of $B$ in **(f)**.

To shed light on the effects of disorder on the WC melting and intermediate state formation, magneto-transport in samples with stronger disorder was compared to clean samples which showed clear intermediate state behavior. Figure 4 shows the 60mK $\rho_{xx}(B)$ of a sample from wafer #1 which is also a 10nm wide QW but with three to six times lower mobility than wafer #5 and #11. At $p=1.38$, this sample shows a peak in $\rho_{xx}(B)$ at ~0.3T, appearing to be similar to wafer #5 and #11 at first sight. However, the temperature dependence of this peak does not show insulating behavior despite the more disordered nature of this sample (Fig.S5). At higher densities, the $\rho_{xx}(B)$ curves consist of regular Shubnikov de Haas (SdH) oscillations without any additional spike-like feature due to the incipient WC formation, in contrast to data from cleaner samples in Fig.2, S1, S2 and Fig.5 below. The practical consequence of stronger disorder in wafer #1 is that for the same resistivity level, one now has higher carrier density or lower interaction parameter $r_s$ and the system is less interacting. For instance, at $\rho_{xx} \sim h/e^2$, $p=0.86$ and $r_s=27.5$ in wafer #5 (Fig.2a) but it requires $p=1.38$ to reach the same level of resistivity (disorder) in wafer #1, corresponding to a lower $r_s=21.7$. The fact that a more disordered sample shows the absence of RIP, and a transport behavior closer to regular Fermi fluid attests to the RIP and intermediate phase formation in clean GaAs QWs with highest $r_s$ being due to Coulomb interaction frustrated 2D WC-liquid transition[10, 15], not a trivial disorder driven puddle formation or inhomogeneity effect.

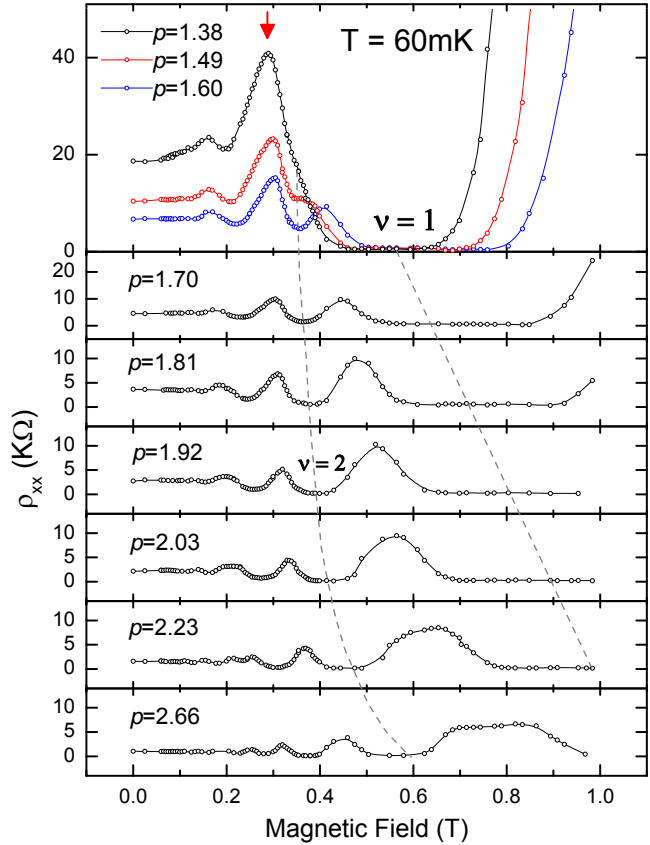

**FIG 4.** Magneto-resistivity of a more disordered QW sample #1-M for density $p$= 1.38-2.66 while current flows along the high mobility direction [$\bar{2}33$]. The red arrow marks the appearance of a RIP like feature at $p$=1.38.

In the various theories of 2D WC melting induced intermediate phase formation [6,7,10,11,12,14], the intermediate phase may have orientation order such as the hexatic liquid crystal [6,14] or stripe phase [10,11] which should exhibit strongly anisotropic transport. The more recent theory of micro-emulsions predicted a number of possible intermediate phases which could be either isotropic or anisotropic: bubbles of a WC in liquid, liquid droplets in a WC, and quasi-one-dimensional stripes in liquid etc[10,11]. We have studied the orientation dependence of the transport in the intermediate phase in our system and excluded the possibility of a quasi-long range ordered stripe phase formation in our experiment. A square Van de Pauw sample was measured in two configurations where the current/voltage probes were rotated by 90 degrees to check any strong anisotropic transport effect due to stripe formation as the intermediate phase. In Figure 5, we observe that the incipient RIP spike appears in both configurations and none of the configurations show a particularly strong insulating behavior at the peak of the RIP spike. The presence of the RIP along both current flow directions was confirmed for multiple densities from $p$=1.43 to 0.92 in this sample (Fig.S1). This indicates that there is no directional ordering of insulating component (i.e. WC) in the intermediate mixture phase over macroscopic length. This is in clear contrast to the striped electronic liquid crystal phase in the second and higher LLs of ultra-high mobility 2D electrons with low $r_s$ values [36, 37]. Our data is consistent with the intermediate phase consisting of WC bubbles in a fluid background. One further intriguing feature is seen in Fig.5 with regard to the effect of extrinsic substrate anisotropy on the transport of intermediate phase. In $B=0$, the corrugation along the $[\bar{2}33]$ direction on (311)A GaAs substrate causes the mobility to be anisotropic and the resistance value for current along the high mobility direction $[\bar{2}33]$ is much smaller than the low mobility direction $[01\bar{1}]$. However, the resistance difference is much smaller in the intermediate phase (marked by the red arrow), suggesting the transport of WC-liquid mixture is less susceptible to extrinsic anisotropic factors.

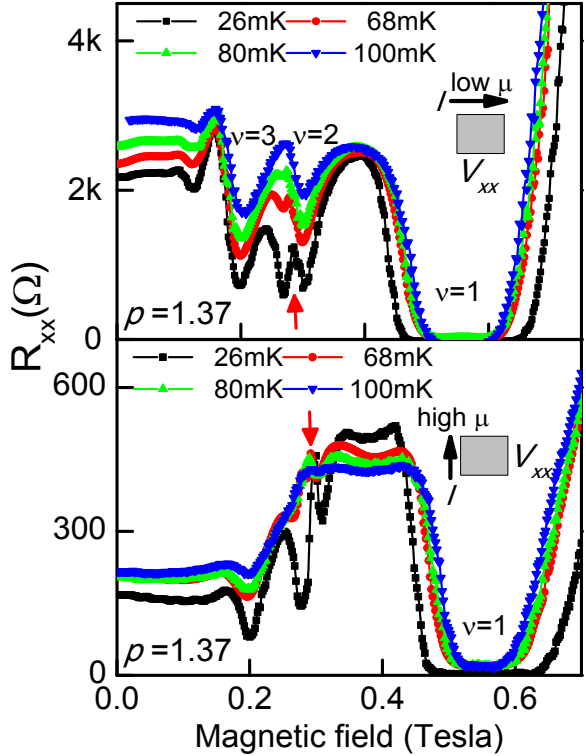

**FIG 5.** (Upper panel) Magneto-resistance of QW #11-J at 55mK, 68mK, 80mK and 100mK for the density $p$= 1.37 while current flows along the low mobility direction $[01\bar{1}]$. (Lower panel) Magneto-resistance measured with current flowing along the high mobility direction $[\bar{2}33]$. In both configurations, the incipient WC formation (marked as red arrows) is apparent, suggesting that the intermediate state of WC and liquid mixture is not a quasi-long range ordered stripe phase.

Coulomb interaction frustrated phase separation has been a prominent subject in many strongly correlated materials. Our observation of an intermediate phase between a Wigner crystal and Fermi liquid in a 2D carrier system in semiconductor incorporates new insights to the understanding of interaction effects on phase transition in low dimensions. Although the presence of intermediate crystal-liquid mixture phase here is clearly revealed via the study of Wigner crystal melting in a small magnetic field, we expect that the physics is relevant in various correlated 2D carrier systems and zero magnetic field as well since the interaction frustrated nature of WC-liquid transition necessarily means that the phase transition is not sharp and intermediate phases exist over broad ranges of the tuning parameters of the transition[15]. More theoretical and experimental investigations regarding the effects of interaction strength, disorder, length scale, and substrate anisotropy would be enlightening to further understand the nature of the intermediate phase observed here and possibly other phases remaining to be uncovered [6-17].

**Acknowledgments** X. P. A. G. thanks NSF for funding support (DMR-0906415). N.G. is partially supported by a US Department of Education GAANN fellowship (grant number P200A090276 and P200A070434). Measurements at the NHMFL High B/T Facility were supported by NSF grant 0654118, by the State of Florida. The work at Princeton was partially funded by the Gordon and Betty Moore Foundation and the NSF MRSEC Program through the Princeton Center for Complex Materials (DMR-0819860).

Supplementary Materials:

Figures S1-S5